\documentclass{sf2a-conf}
\usepackage{graphicx}
\usepackage{natbib}
\usepackage{multirow}

\def\BLF{bulk Lorentz factor }
\def\BDF{Doppler factor }
\def\pks{PKS~2155--304 }

\newcommand{\apj}{ApJ}
\newcommand{\apjl}{ApJL}
\newcommand{\mnras}{MNRAS}
\newcommand{\aap}{A\&A}

\newcommand{\pasp}{PASP}

\begin{document}

\TitreGlobal{SF2A 2008}

\title{Time dependent modelisation of TeV blazars by a stratified jet model}
\author{Boutelier, T.}\address{Laboratoire d'Astrophysique de Grenoble--Universit\'eŽ Joseph-Fourier/CNRS UMR 5571 --BP~53, F-38041 Grenoble, France}
\author{Henri, G. $^1$}
\author{Petrucci, P-O.$^1$}
\runningtitle{Time dependent modelisation of TeV blazars}
\setcounter{page}{237}

\index{Boutelier, T.}
\index{Henri, G.}
\index{Petrucci, P.-O.}

\maketitle
\begin{abstract} 
We present a new time-dependent inhomogeneous jet model of non-thermal blazar emission. Ultra-relativistic leptons are injected at the base of a jet and propagate along it. We assume continuous reacceleration and cooling, producing a relativistic quasi-maxwellian (or "pile-up") particle energy distribution. The synchrotron and Synchrotron-Self Compton jet emissivity are computed at each altitude. Klein-Nishina effects as well as intrinsic gamma-gamma absorption are included in the computation. Due to the pair production optical depth, considerable particle density enhancement can occur, particularly during flaring states.Time-dependent jet emission can be computed by varying the particle injection, but due to the sensitivity of pair production process, only small variations of the injected density are required during the flares. 
The stratification of the jet emission, together with a pile-up distribution, allows significantly lower bulk Lorentz factors, compared to one-zone models. Applying this model to the case of \pks and its big TeV flare observed in 2006, we can reproduce {\it simultaneously} the average broad band spectrum of this source from radio to TeV, as well as TeV light curve of the flare  with  \BLF lower than 15.
\end{abstract}
%
\section{Introduction}\label{intro}
It is widely admitted that the blazar phenomenon is due to relativistic Doppler boosting of the non-thermal jet emission taking place in radio-loud Active Galactic Nuclei (AGN) whose jet axis is closely aligned with the observer's line of sight. Blazars exhibit very broad spectral energy distributions (SED) ranging from the radio to the gamma-ray band and dominated by two broad band components. In the Synchrotron Self Compton scenario (SSC), the lowest energy hump is attributed to the synchrotron emission of relativistic leptonic particles, and the highest one is attributed to the Inverse Compton process (IC) of the same leptons on the synchrotron photon field. Broad band observations of these objects are crucial to understand the jet physics and to put reliable constraints on jet parameters.

The most extreme class of blazars are the highly peaked BL lac sources (HBL), where the synchrotron/Inverse Compton components peak in the UV/X-ray/gamma-ray (GeV up to TeV)  range. These objects are well known to be highly variable in all energy bands. Perhaps the most extreme example of this extraordinary variability behaviour has been caught by the HESS instrument with the big flare of \pks during summer 2006 (Aharonian et al. 2007).

In this case,  the observed variability time scale ($\sim$ 200 sec) in the TeV range implies a minimum \BLF greater than 50 (Begelman, Fabian, \& Rees 2008) assuming an homogeneous one zone model. However, such high values of the \BLF are in contradiction with constrains derived from other observational evidence (Urry \& Padovani 1995; Henri \& Saug\'e 2006 and references therein). Furthermore, one-zone models are unable to fit the entire spectrum, the low energy radio points being generally attributed to more distant emitting regions.  
We present here a new approach, unifying small and large scales emission regions:  we consider that the radio jet is actually filled by the same particles originating from the high energy emitting region, at the bottom of the jet, that have propagated along it. We describe thus the emitting plasma by a continuous (although variable) particle injection, submitted to continuous reacceleration and radiative cooling. This model fits well into the two-flow framework originally proposed by Pelletier (1985) and Sol et al. (1989) where a non relativistic, but powerful MHD jet launched by the accretion disk, surrounds a highly relativistic plasma of electron-positron pairs propagating along its axis. The MHD jet plays the role of a collimater and an energy reservoir for the pair plasma, which is responsible for the observed broad band emission. 

\section{Description of the model}\label{sec:model}
\subsection{Geometry of the model and particle energy distribution}

\begin{figure}
\centering
\parbox{0.5\linewidth}{
\includegraphics[angle=90,width=1.0\linewidth]{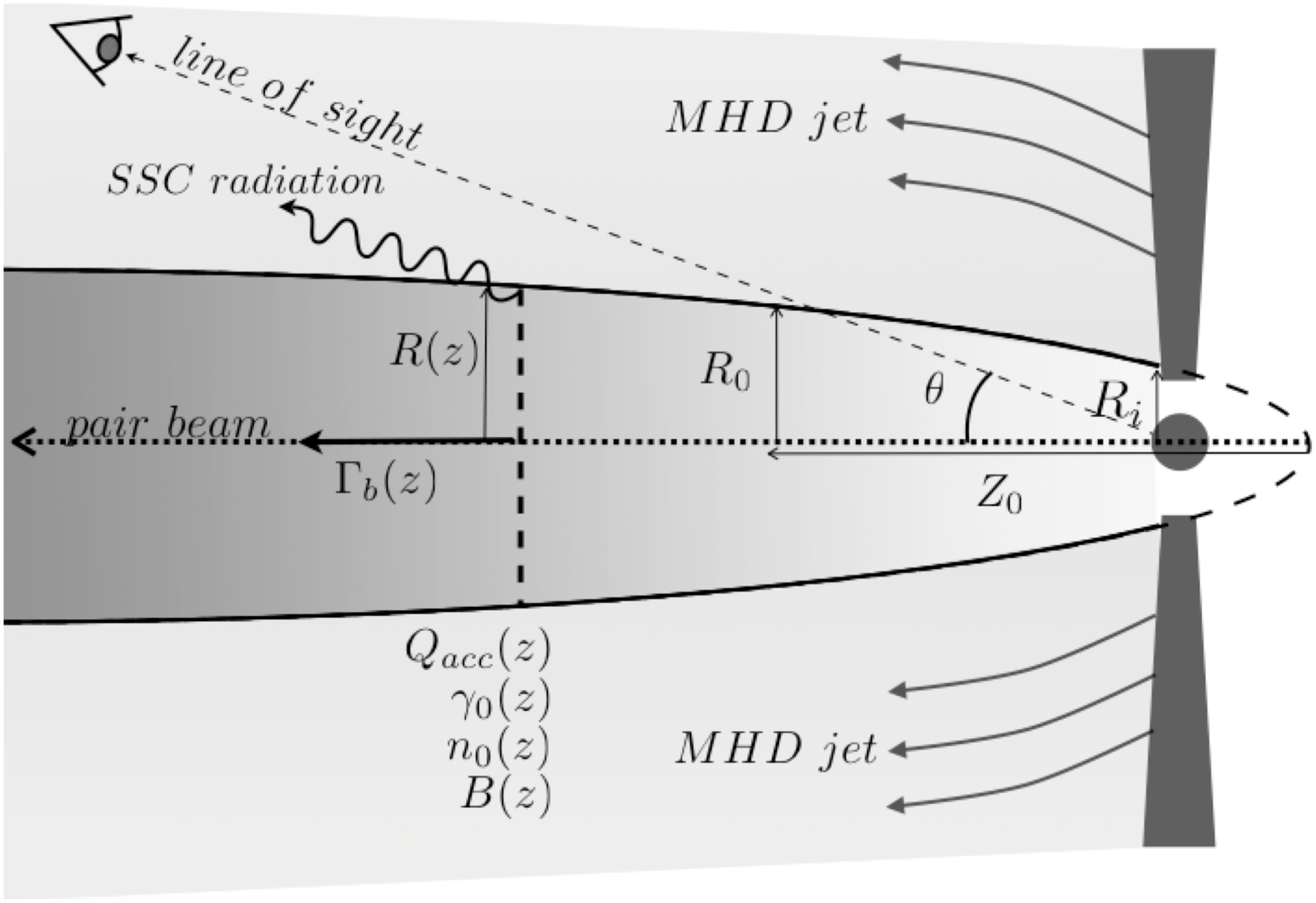}
}
\quad
\parbox{0.45\linewidth}{
\caption{\label{fig:geo} {Sketch of the jet geometry. See text for the signification of the different parameters.}}
}
\end{figure}


We consider  that the relativistic plasma propagates in a stationary funnel (associated to the last magnetic surface of the surrounding MHD jet) that we parametrize as a shifted paraboloid shape (power-law with an index $\omega$). The magnetic field scales with the jet radius to the power $-\lambda$. We consider that the jet is continuously accelerating, starting from rest ($\Gamma_b = 1$ at $z=0$) and reaching an asymptotic value $\Gamma_{b \infty}$ on a typical scale $Z_{0}$. 

The MHD jet that surrounds the radiative pair plasma transfers its energy to the pairs via second order Fermi acceleration mechanism (particle-wave interaction). As a result of this acceleration mechanism, the energy distribution function (EDF) of the electron-positron plasma is assumed to be a relativistic maxwellian (or ``pile-up'') distribution (Saug\'e \& Henri 2004 and references therein).
As they propagate inside the structure, the particles lose energy via synchrotron and inverse Compton cooling. Particles are continuously re-heated by the surrounding MHD jet, via a Fermi II acceleration process. The acceleration rate $Q_{acc}(z)$ is parametrized as a power-law with an index $-\zeta$ and a normalisation factor $Q_{0}$. To avoid energy divergence, the heating is stopped after an altitude $Z_{c}$. The relativistic maxwellian's ``temperature'' $\gamma_{0}$ is determined by balancing radiative cooling and re-heating. Because of absorption of $\gamma$-ray photons via the pair creation process, new particles are created inside the jet, which yields to the increase of the particle flux ${\Phi (z,t) = \int n(\gamma,z,t) S(z) \Gamma_b \beta_b c d\gamma}$, where $S(z)$ is the surface of radial section of the jet. This quantity is computed via a continuity equation that takes into account the pair creation term. Fig. 1. displays the different parameters used in the model. More details about the parametrisation we use can be found in Boutelier et al. (2008).

 \subsection{The jet spectrum}
We compute the emissivity at each altitude in the jet by assuming a SSC process for the radiative mechanism. The total intensity of the jet is then determined by integrating the emissivity all along the jet.
Once injected at the base of  the jet, the particles contribute first to the high energy part of the jet SED. As they propagate, their emissivity peaks progressively  at lower energy, producing the low energy part of the spectrum. The spectral shape of the whole SED is not controlled by the local particle energy distribution (which is always locally a narrow pile-up), but rather by the z-dependencies of the jet radius, the magnetic field, and the acceleration rate. 
A constant injection rate would lead to a stationary emission pattern,  which would be rather easy to fit. In reality, the observed instantaneous spectra are a complicated convolution of the whole history of the jet, keeping the memory of the whole past injection pattern. Boutelier et al. (2008) describes a procedure to extract the physical parameters of the jet from observed spectra, despite the fact that they do not correspond to a simple steady-state of the jet.


\section{Application to \pks}\label{sec:apli}
\begin{table*}
\caption{Model parameters of the flaring and quiescent state. During the flaring state, the flux of injected particles varies between the indicated minimum and maximum values, following the injection pattern displayed in Fig. 2. The other parameters remain fixed. $R_{i}$, $R_{0}$, $Z_{0}$, $Z_{c}$ are in unit of $10^{14}cm$ and are displayed on Fig. 1.}
\label{tab:param}
\begin{center}
\begin{tabular}{cccccccccccccc}
\hline
STATE && $\Phi(Z_i)$ & $\Phi(Z_0)$ & $Q_0$ &$\Gamma_{b \infty} $ & $R_i$&$R_0$ & $Z_0$ & $Z_c$ & $B_0$ & $\omega$ & $\lambda$& $\zeta$  \\
  & & $[10^{42} s^{-1}]$& $[10^{42} s^{-1}]$ &$[s^{-1}]$& & $$ &  &  &  & $[G]$ &  &  &    \\  
\hline
\multirow{3}*{ flaring} & Max&$2.09 $ & $70.1 $& \multirow{3}*{ $6.5$}\\ 
 & Aver.&$1.84 $& $24.4$ &\\ 
  & Min&$1.16 $& $2.33 $&  \\  \cline{1-5}
{ quiescent} & &$1.16$ &$1.55$&$2.5$ & \multirow{-2}*{15} & \multirow{-2}*{$1.1$}& \multirow{-2}*{$1.78$}&\multirow{-2}*{$20$}& \multirow{-2}*{$5\times 10^{7}$}&\multirow{-2}*{5}&\multirow{-2}*{0.2}&\multirow{-2}*{1.9}&\multirow{-2}*{1.27} \\  
\hline
\end{tabular}
\end{center}
\end{table*}

\begin{figure}[ht]
\begin{center}
        \resizebox{9cm}{!}{\includegraphics  {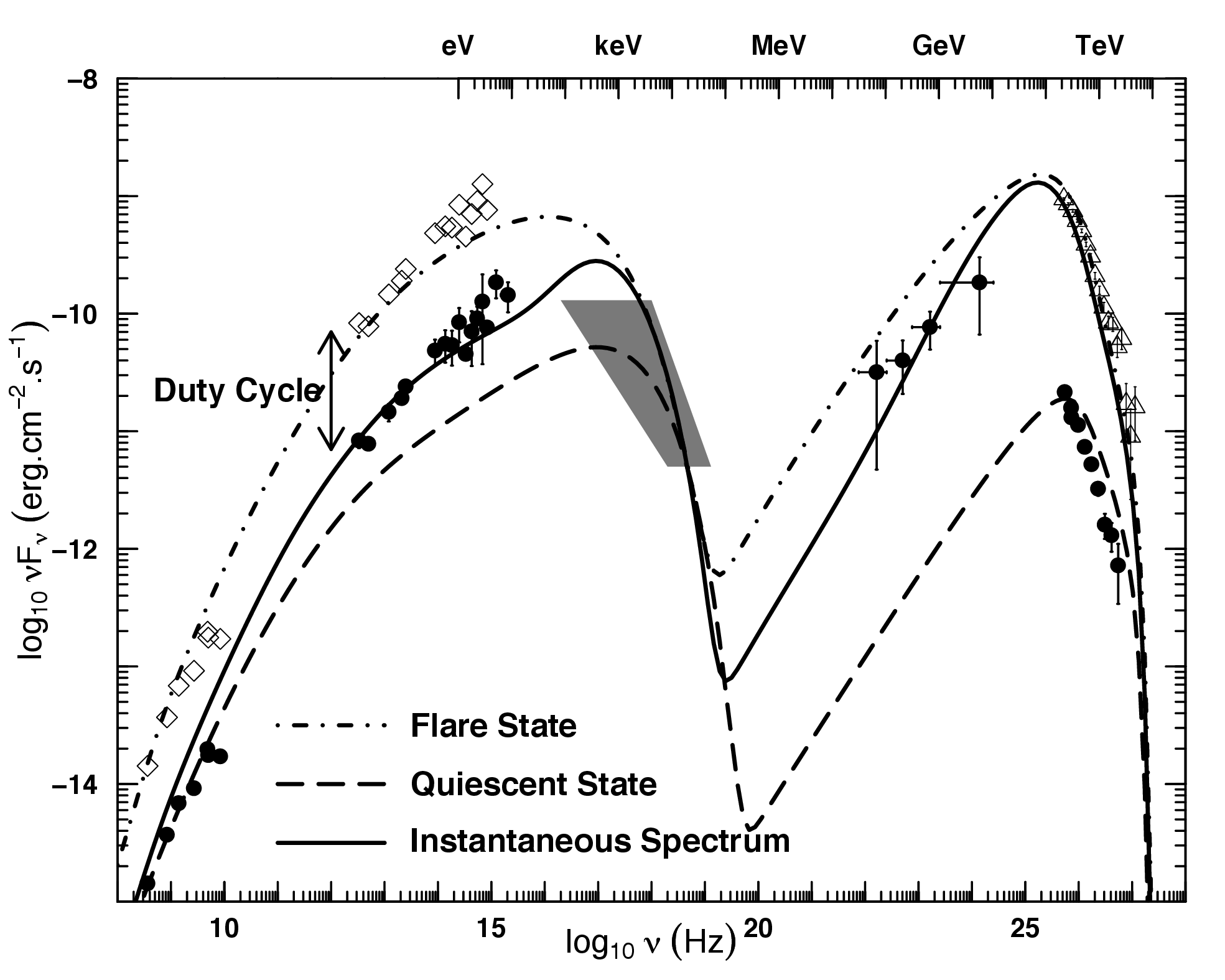}}
        \resizebox{7cm}{!}{\includegraphics  {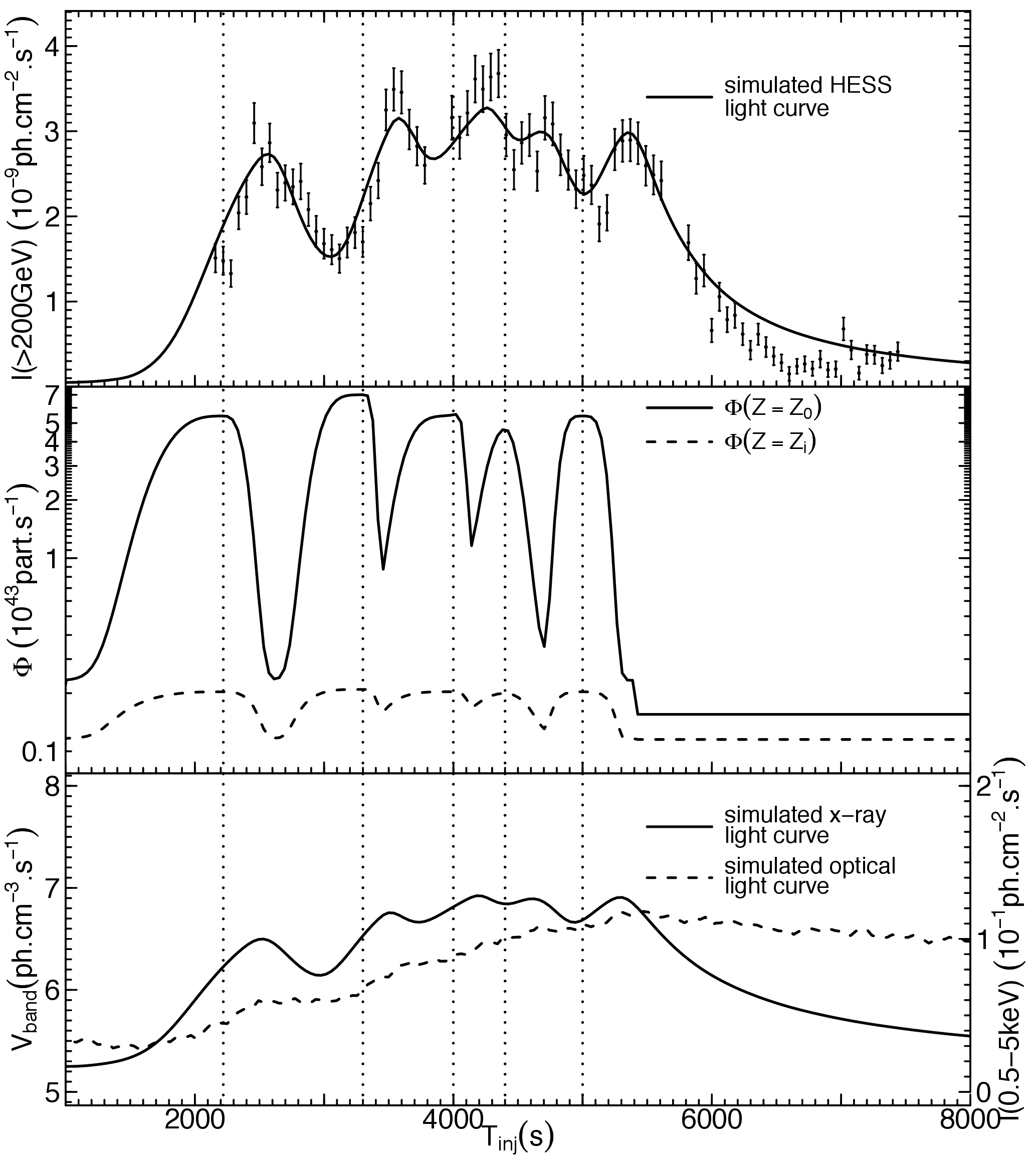}}
\caption{{\bf Left} : Fit of \pks data. {\bf Filled dots:} average archival data (see text). {\bf Empty triangles:} average HESS data from the big flaring night. {\bf Empty diamonds:} "fake flaring" state low energy points {{\bf Shaded area:} enveloppe of archival X-ray data from BeppoSAX, SWIFT and XMM-Newton}. {\bf Dot-dashed line:} best fit of the " average fake flaring" spectrum. {\bf Dashed line:} fit of the { quiescent ($\sim$average)} spectrum. {\bf Solid line:} example of an instantaneous  {\bf simulated spectrum}. 
{\bf Right}:  {\bf Upper panel}: HESS light curve above 200~GeV superimposed with the model (solid line). {\bf Middle panel}: time dependent particle injection function used in the simulation. {\bf Lower panel}: predicted light curves in X-ray (dashed line, left y-scale) and optical (dot-dashed line, right y-scale). The dotted lines mark the maximum of the different bursts of the injection function.
}
\label{fig2}
\end{center}
\end{figure}

We have tested our model to  the big flare event observed by HESS in July 2006 in \pks. We apply the method described in Boutelier et al. (2008) to the SED of PKS~2155--304.  The average spectrum of \pks in the low frequencies range ($\le10^{15}~Hz$), is constructed using archival data from the HEASARC archive website (http://heasarc.gsfc.nasa.gov/docs/archive.html). Based on the activity detected by HESS, we estimate that the duty-cycle is around 10\%. Then to construct the "fake flaring" spectrum of this source, i.e. the spectrum expected if the jet was always in a flaring state, we combine the HESS data with the radio-to-optical ones corrected by a factor 10 (dot-dashed line in left plot of Fig. 2).

An average "fake flaring" spectrum is shown in the left side of Fig.~\ref{fig2} in dot-dashed line. The corresponding best fit model parameters are reported in Tab.~\ref{tab:param}. Interestingly we only need a \BDF $\Gamma_b=15$ which is significantly below the values of $\sim$ 50 inferred from one-zone model (Begelman, Fabian, \& Rees 2008).
The best fit parameters of the TeV quiescent state (derived from averaged H.E.S.S spectrum in Aharonian et al. 2005) are also reported in Tab.~\ref{tab:param} and the solution has been overplotted in the left side of Fig.~\ref{fig2} (dashed line).  

%
To reproduce the observed H.E.S.S. light curve during the big-flare event, we use an injection function $\Phi(z=0,t)$ that is the sum of five "generalized Gaussian" shape, similar to the analysis made in Aharonian et al. (2007).
Before the flare, the flux of particles $\Phi(z=0,t)$ is assumed to be a crenel function that oscillates between quiescent and flaring states in agreement with the source duty cycle.
At the right top of  Fig. \ref{fig2}, we have reported the HESS light curve and the simulated one (solid line). The agreement between the simulated and the reported HESS light curve is very good. The instantaneous spectrum overplotted  in solide lines on the left side of Fig. \ref{fig2} agrees with the radio to TeV spectrum observed during the burst.

\section{Discussion}\label{sec:disc}

Our time-dependent inhomogeneous jet model succeeds in  reproducing {\it simultaneously} the broad band (from radio to TeV) spectrum of \pks as well as the TeV light curve during the big flare event of July 2006. The key idea of the method is to decompose the blazar spectrum in "quiescent", low luminosity states, and "flaring", high luminosity states. The high energy part of the spectrum, coming from  small-scale inner regions, is assumed to be, at any time, in one of these pure states. On the other hand the low energy part is a convolution over a large scale of the past history of the jet : it is thus rather a time-averaged spectral state mixing quiescent and flaring states in proportions given by the source "duty-cycle". Moreover we do not require two different populations of emitting particles like in other models (e.g. Katarzy\'nski et al.~2003) but simply a continuous (although variable) injection of particles at the base of the jet that propagate along the jet structure. Pair production plays an important role to amplify the initial variation, as can be seen with the variation of the particle flux along the jet during the flaring state (see Tab. 1 and the right side of Fig. 2 (middle)): the pair current is amplified by a factor 30 at the end of the jet, while the initial current varies only by a factor 2. 

The model can also predict light curves at different wavelengths. As an example, the x-ray (2-10 keV) and optical (V band) light curves expected during the TeV flare have been plotted at the right bottom of Fig.\ref{fig2}. The x-ray luminosity exhibits almost simultaneous variations but with a lower amplitude (about 5 times smaller). On the other hand, the optical light curve shows a very different behavior, increasing all along the flare. This is due to to the large size of the optical emitting region that plays the role of a low pass filter. Consequently, the optical luminosity integrates the recent past history of the jet.
These results are compatible with simultaneous multiwavelength observations made during the ``Chandra night'' (Costamante, private communication).

In the present work, we do not specify the origin of the variability. Obviously, other plasma parameters can vary in addition to the injection density $N_0$. Most likely, variability  can be triggered by a change in the acceleration rate described by $Q_{acc}$. In a plausible scenario, long term (year scale) variability implying the succession of quiescent and active states could be attributed to variations in the accretion rate, wheras the short (minute-scale) flares would be attributed to the instability to pair creation that develops only when the initial particle density is close to a critical threshold.



\end{document}